\documentclass[conference]{IEEEtran}

\usepackage{cite}
\usepackage{color}
\usepackage[dvips]{graphicx}
\usepackage[cmex10]{amsmath}
\usepackage{psfrag}
\usepackage{bm}
\usepackage{comment}

\def \bmx {\bm{x}}
\def \bmy {\bm{y}}

\def \bmh {\bm{h}}
\def \bmz {\bm{z}}
\def \bmv {\bm{v}}
\def \EE  {\mathsf{E}}
\def \llangle     {\langle \! \langle}
\def \rrangle     {\rangle \! \rangle}
\def \llanglebig  {\big \langle \! \big \langle}
\def \rranglebig  {\big \rangle \! \big \rangle}
\def \llangleBig {\Big \langle \!\! \Big \langle}
\def \rrangleBig  {\Big \rangle \!\! \Big \rangle}
\def \llangleBigg {\Bigg\langle\kern-4.2pt\Bigg\langle}
\def \rrangleBigg {\Bigg\rangle\kern-4.2pt\Bigg\rangle}
\def \Ltwo {\big|\!\big|}

\newtheorem{claim}{Claim}

\begin{document}

\title{Microscopic Analysis for Decoupling Principle of Linear Vector
Channel}
\author{\IEEEauthorblockN{Kazutaka Nakamura and Toshiyuki Tanaka}
\IEEEauthorblockA{Department of Systems Science, Graduate School of
Informatics, Kyoto University\\
36-1, Yoshida-Honmachi, Sakyo-ku, Kyoto, 606-8501, Japan\\
Email: \{knakamur,~tt\}@i.kyoto-u.ac.jp}}
\maketitle

\begin{abstract}
This paper studies the decoupling principle of a linear vector
channel, which is an extension of CDMA and MIMO channels. 
We show that the scalar-channel characterization 
obtained via the decoupling principle is valid 
not only for collections of a large number of elements of input vector, 
as discussed in previous studies, 
but also for individual elements of input vector,
i.e. the linear vector channel for individual elements of channel input vector
is decomposed into a bank of independent scalar Gaussian channels
in the large-system limit, where 
dimensions of channel input and output are both sent to 
infinity while their ratio fixed.
\end{abstract}

\IEEEpeerreviewmaketitle

\section{Introduction}
Recently, the replica method, developed in statistical mechanics, 
has been applied to problems of performance evaluation 
of various digital wireless communication systems,
especially code-division multiple-access (CDMA)
and multi-input multi-output (MIMO)
systems~\cite{ART__TT_IEEE_Trans_IT_2002,ART__DG_SV_IEEE_Trans_IT_2005,ART__RRM_2003,INP__KT_TT_2007}. 
The replica method provides us with a description of these channels, 
called, the {\em decoupling principle}; 
that is, a CDMA channel, or equivalently a MIMO channel, 
is decoupled, under a certain randomness assumption of the channel, 
into a bank of independent scalar Gaussian channels 
in the large-system limit, where 
dimensions of channel input and output are both sent to 
infinity while their ratio fixed.  

Existing results of replica analysis, however, 
rely on saddle-point evaluation of integrals, 
which is only valid for evaluating {\em macroscopic} quantities, 
such as an empirical mean of many {\em microscopic} quantities, 
such as individual elements of input, 
which are many in the sense that their number goes to infinity 
as the dimensions of the system in the large-system limit. 
It is therefore not clear as to whether the scalar-channel 
characterization of CDMA or MIMO channels obtained 
via the replica analysis is still valid 
if we are interested in microscopic quantities 
in the large-system limit. 

In this paper we show that 
the scalar-channel characterization is still valid for 
microscopic quantities, 
by performing replica analysis on a linear vector channel, 
which is an extension of CDMA or MIMO channels.

\section{Linear Vector Channel}
We consider a $K$-input $N$-output linear vector channel, defined as follows.
Let $\bmx_0=(x_{01},\,\ldots,\,x_{0K})^T$ denote the input vector of the channel, 
and $\bmy=(y_1,\,\ldots,\,y_N)^T$ denote the output vector 
given a linear transform $H\bm{x}_0$ of the inputs, 
where $H$ is an $N \times K$ channel matrix.  
Assuming the channel to be memoryless, 
the input-output characteristic of the linear
vector channel is represented as 
\begin{align}
P_0(\bmy|H \bmx_0 )=\prod_{\mu=1}^N \rho_0
\left(y_\mu \Bigg| \frac{\bmh_\mu^T \bmx_0}{\sqrt{N}} \right),
\label{eq:true_vector_channel}
\end{align}
where $\bmh_\mu^T/\sqrt{N}$ denotes $\mu$\,th row of $H$.
We define a true prior as $P_0(\bmx)$.
Inference of the input vector $\bmx_0$, given the output vector $\bmy$ 
and the channel matrix $H$, can be solved by 
a detection scheme based on Bayesian inference.
The detector assumes a channel model to be 
$P(\bmy|H\bmx)=\prod_{\mu=1}^N \rho(y_\mu|\bmh_\mu^T \bmx/\sqrt{N})$, 
and a prior distribution to be $P(\bmx)$. 
We also assume perfect channel state information at the detector.  
These assumptions yield the posterior distribution 
\begin{align}
P(\bmx|\bmy,\,H)=
\frac{P(\bmy|H\bmx)P(\bmx)}{\int P(\bmy|H\bmx)P(\bmx)\,d\bmx }.
\label{eq:Posterior_Prob}
\end{align}
The posterior mean estimator (PME) 
$\bar{\bmx} = \int \bmx \,P(\bmx|\bmy,\,H)\,d\bmx$ 
is the optimal inference scheme to minimize the mean squared error, 
if the assumed model is matched to the true model. 

In this paper, 
we study joint distributions of $L$ ($\ll K$) elements of input vector 
and their estimates based on the posterior
distribution~\eqref{eq:Posterior_Prob}, given a channel matrix $H$.  
Without loss of generality we consider the first $L$ elements 
of input vector, $\bmx_0^L=(x_{01},\,\ldots,\,x_{0L})^T$,  
and their estimates $\bmx^L$.  
The joint distribution to be studied is thus 
\begin{align}
\mathcal{P}(\bmx_0^L,\,\bmx^L|H) =
\int 
P(\bmx^L|\bmy,\, H)
 P_0(\bmy|H \bmx_0)P_0(\bmx_0)
\,d\bmx_0^{\backslash L}.
\end{align}
We assume the channel matrix $H$ to be random and 
evaluate expectation of $\mathcal{P}(\bmx_0^L,\,\bmx^L|H)$ 
over $H$ in the large-system
limit where $K$, $N \to \infty$ while $\beta = K/N$ is kept finite: 
\begin{align}
\mathcal{P}(\bmx_0^L,\,\bmx^L) 
= \lim_{K,\,N\to\infty}\EE_H \left[
\mathcal{P}(\bmx_0^L,\,\bmx^L|H)
\right].
\label{eq:joint_distt}
\end{align}
$\EE_u[\cdots]$ denotes expectation over the random
variable $u$.
Note that if the scalar-channel characterization is derived for the 
joint distribution \eqref{eq:joint_distt} using the replica method, 
it is easy to show the scalar-channel characterization is still valid for 
arbitrary microscopic quantities depend on $\bmx_0^L$ and $\bmx^L$.

To simplify the analysis, we assume the following:
\begin{itemize}
\item Random channel matrix: The elements $\{h_{\mu k}\}$ 
are independent and identically distributed (i.i.d.) with mean zero,
unit variance, odd-order moments being zero and $(2m)$\,th-order moments
being finite.
\item The first $L$ elements of input vector $\bmx_0^L$ 
 and the remaining elements $\bmx_0^{\backslash L}=(x_{0(L+1)},
\,\ldots,\, x_{0K})^T$ are independent, so that 
the prior distribution of $\bmx$ is factorized as
\begin{align}
P_0(\bmx_0)= P_0^L (\bmx_0^L) \,
P_0^{\backslash L}(\bmx_0^{\backslash L}).
\end{align}
The factorized form $P(\bmx) = P^L (\bmx^L) \, P^{\backslash L}
(\bmx^{\backslash L})$ is also used as the postulated prior distribution. 
\item The conditional distributions
$\rho_0(y|u)$ and $\rho(y|u)$ are one and two times differentiable with
respect to $u$, respectively.
\end{itemize}

\section{Main Result}
Our main result is the following claim.  

\begin{claim}
In the large-system limit and under the assumption of replica symmetry 
(see Sect.~\ref{sec:derivation}),
the joint distribution
$\mathcal{P}(\bmx_0^L,\,\bmx^L)$ defined in \eqref{eq:joint_distt}
is asymptotically equivalent to the joint distribution
\begin{align}
\mathcal{P}(\bmx_0^L,\,\bmx^L)
&=
\int
\frac{\prod_{k=1}^L \rho_G(z_k|x_k) \, \tilde{P}^L(\bmx^L)}
{\int \prod_{k=1}^L \rho_G(z_k|x_k) \, \tilde{P}^L(\bmx^L) \, d\bmx^L}
\nonumber \\
&\times \prod_{k=1}^L
\rho_{G0}(z_k|x_{0k}) \, P_0(\bmx_0^L) \,d\bmz^L,
\label{eq:decoupled_joint_distt}
\end{align}
where
$\rho_{G0}(z|x)$ and $\rho_{G}(z|x)$ represent input-output characteristics 
of the scalar Gaussian channels 
\begin{align}
\rho_{G0}(z|x) &= \sqrt{\frac{E^2}{2 \pi F}} \exp \left[
-\frac{E^2(z-x)^2}{2 F}
\right],
\\
\rho_{G}(z|x) &= \sqrt{\frac{E}{2 \pi}} \exp \left[
-\frac{E(z-x)^2}{2}
\right],
\end{align}
respectively, 
and where $\bmz^L=(z_1,\,\ldots,\,z_L)^T$.
$\tilde{P}^L(\bmx^L)$ is a ``modulated'' version of the assumed prior, 
defined as 
\begin{align}
\tilde{P}^L(\bmx^L)=
\frac{\exp \Big[\frac{G-F+E}{2}\,\Ltwo \bmx^L \Ltwo^2 \Big] P^L(\bmx^L)}
{\int \exp \Big[\frac{G-F+E}{2}\,\Ltwo \bmx^L \Ltwo^2 \Big]
P^L(\bmx^L)\,d\bmx^L},
\end{align}
where $\Ltwo \bm{x} \Ltwo^2=\bm{x}^T\bm{x}$.

The parameters $\{G,\,E,\,F\}$
are determined by solving the
following equations for $\{G,\,E,\,F,\,r,\,m,\,q\}$, 
\begin{align}
G &= \int \bar{\rho}_0 \left(
y\Bigg|\sqrt{\frac{\beta m^2}{q}\,t}
\right)
\frac{\bar{\rho}^{\prime\prime}\left(y|\sqrt{\beta q}\,t\right)}
{\bar{\rho}\left(y|\sqrt{\beta q}\,t\right)}
\,Dt\,dy,
\label{eq:SaddlePoint_G}
\\
E&= \int \bar{\rho}_0^\prime \left(
y\Bigg|\sqrt{\frac{\beta m^2}{q}\,t}
\right)
\frac{\bar{\rho}^\prime \left(y|\sqrt{\beta q}\,t\right)}
{\bar{\rho}\left(y|\sqrt{\beta q}\,t\right)}
\,Dt\,dy,\\
F&= \int \bar{\rho}_0 \left(
y\Bigg|\sqrt{\frac{\beta m^2}{q}\,t}
\right)
\left[\frac{\bar{\rho}^\prime \left(y|\sqrt{\beta q}\,t\right)}
{\bar{\rho}\left(y|\sqrt{\beta q}\,t\right)}
\right]^2
\,Dt\,dy,
\\
r &= \lim_{K \to \infty} \frac{1}{K}
\llangleBig
\Ltwo \langle \bmx \rangle \Ltwo^2
\rrangleBig, \\
m &= \lim_{K \to \infty} \frac{1}{K}
\llangleBig \bmx_0^T \langle \bmx \rangle \rrangleBig,
\\
q &= \lim_{K \to \infty} \frac{1}{K}
\llangleBig \Ltwo \langle \bmx \rangle \Ltwo^2 \rrangleBig,
\label{eq:SaddlePoint_q}
\end{align}
where 
$\int (\cdots)\,Du = \int_{-\infty}^\infty
(\cdots)\exp(-u^2/2)\,du/\sqrt{2\pi}$. 
The distributions $\bar{\rho}_0$ and $\bar{\rho}$ are defined as 
\begin{align}
&\bar{\rho}_0\left(
y \Bigg| \sqrt{\frac{\beta m^2}{q}}\,t \right)
\nonumber \\
&= \int \rho_0 \left(
y \Bigg| \sqrt{\frac{\beta m^2}{q}}\,t
+ \sqrt{\beta \left( r_0 - \frac{m^2}{q} \right)}\, u
\right)
\,Du,
\\
&\bar{\rho}\left(
y\Big|\sqrt{\beta q}\,t \right)
= \int \rho \left(
y\Big|\sqrt{\beta q}\,t
+ \sqrt{\beta \left( r - q \right)}\, u
\right)
\,Du,
\end{align}
respectively, 
where 
$f^\prime(y|u) = \frac{\partial}{\partial u} f(y|u)$, 
and where 
\begin{equation}
r_0 = \lim_{K \to \infty}
\frac{1}{K}
\int \Ltwo \bmx_0 \Ltwo^2 P_0(\bmx_0)
\,d\bmx_0.
\label{eq:SaddlePoint_r0}
\end{equation}
The brackets $\llangle \cdots \rrangle$ and $\langle \cdots \rangle$
denote the averages with respect to the joint distribution of
$\bmx_0$ and $\bmz=(z_1,\,\ldots,\,z_K)^T$, 
\begin{equation}
\llanglebig \cdots
\rranglebig
=
\iint (\cdots) \prod_{k=1}^K \
\rho_{G0}(z_k|x_{0k})
\,P_0(\bmx_0)\, d \bmz \, d\bmx_0,
\end{equation}
and the posterior distribution of $\bmx$ given~$\bmz$,
\begin{equation}
\langle \cdots \rangle
=
\frac{\int (\cdots) \prod_{k=1}^K
\rho_{G}(z_k|x_k) \,\tilde{P}(\bmx)}
{\int \prod_{k=1}^K
\rho_{G}(z_k|x_k) \,\tilde{P}(\bmx)\,d\bmx},
\end{equation}
respectively, where
%
\begin{align}
\tilde{P}(\bmx) &=
\frac{\exp \Big[ \frac{G-F+E}{2}\Ltwo \bmx \Ltwo^2 \Big] P(\bmx)}
{\int \exp \Big[ \frac{G-F+E}{2}\Ltwo \bmx \Ltwo^2 \Big] P(\bmx)\,d\bmx}.
\end{align}

If more than one solution exists for
\eqref{eq:SaddlePoint_G}--\eqref{eq:SaddlePoint_q}, the correct solution is
the one that minimizes the function $\mathcal{F}$ defined as 
\begin{align}
\mathcal{F} &=
\frac{1}{\beta} \iint \bar{\rho}_0\left(y\Bigg|\sqrt{\frac{\beta m^2}{q}}\,t
\right)
\log \bar{\rho}\left( y \Big|\sqrt{\beta q}\,t
\right)
\,Dt\,dy
\nonumber \\
&+\frac{1}{2} G r - E m + \frac{1}{2} F q
+ \frac{F}{2E} + \frac{1}{2}E r_0 - \frac{1}{2}\log \frac{E}{2 \pi}
\nonumber \\
& + \lim_{K \to \infty} \frac{1}{K} \iint \prod_{k=1}^K
\rho_{G0}(z_k|x_{0k})\,P_0(\bmx_0)
\nonumber \\
&\times \left\{ \log \int \prod_{k=1}^K
\rho_{G}(z_k|x_k) \, \tilde{P}(\bmx)\,d\bmx
\right\}
\,d\bmx_0
\,d\bmz.
\label{eq:free_energy}
\end{align}
\label{claim}
\end{claim}

\begin{figure}
\psfrag{b0_sim_P0b0}{$\bmx_0 \sim P_0(\bmx_0)$}
\psfrag{b01}{$x_{01}$}
\psfrag{b02}{$x_{02}$}
\psfrag{b0L}{$x_{0L}$}
\psfrag{b0K}{$x_{0K}$}
\psfrag{b1}{${\scriptstyle x_1}$}
\psfrag{b2}{${\scriptstyle x_2}$}
\psfrag{bL}{${\scriptstyle x_L}$}
\psfrag{bK}{${\scriptstyle x_K}$}
\psfrag{hatb1}{$\hat{x}_1$}
\psfrag{hatb2}{$\hat{x}_2$}
\psfrag{hatbL}{$\hat{x}_L$}
\psfrag{hatbK}{$\hat{x}_K$}
\psfrag{P0_CDMA_channel}{$P_0(\bmy|H \bmx_0)$}
\psfrag{sy}{${\scriptstyle \bmy}$}
\psfrag{y}{$\bmy$}
\psfrag{post_prob_true}{${\scriptstyle P(\bmx|\bmy,H)}$}
\psfrag{pb}{${\scriptstyle P(\bmx)}$}
\psfrag{p_channel_true}{${\scriptstyle P(\bmy|H\bmx)}$}
\begin{center}
\includegraphics[scale=1.0]{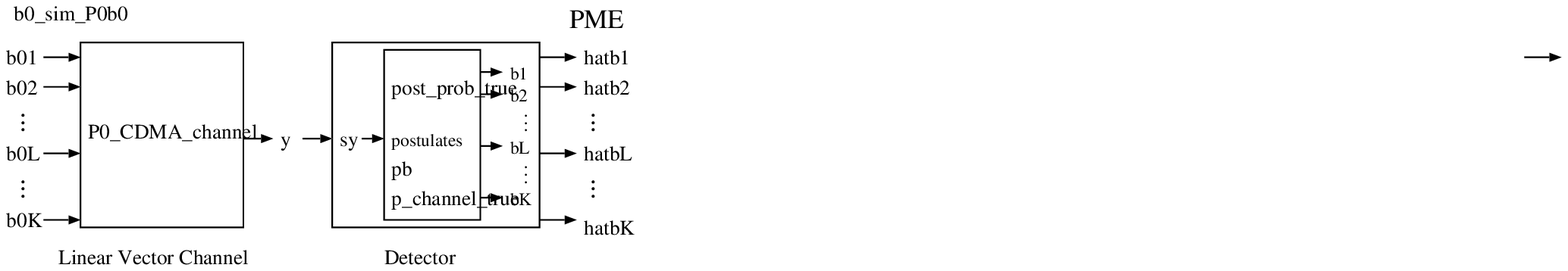}
(a) \vspace{-5mm}
\end{center}
\psfrag{b0L_sim_P0Lb0L}{$\bmx_0^L \sim P^L_0(\bmx_0^L)$}
\psfrag{rho_G0_1}{$\rho_{G0}(z_1|x_{01})$}
\psfrag{rho_G0_2}{$\rho_{G0}(z_2|x_{02})$}
\psfrag{rho_G0_L}{$\rho_{G0}(z_L|x_{0L})$}
\psfrag{rho_G_1}{${\scriptstyle \rho_G(z_1|x_1)}$}
\psfrag{rho_G_2}{${\scriptstyle \rho_G(z_2|x_2)}$}
\psfrag{rho_G_L}{${\scriptstyle \rho_G(z_L|x_L)}$}
\psfrag{sz1}{${\scriptstyle z_1}$}
\psfrag{sz2}{${\scriptstyle z_2}$}
\psfrag{szL}{${\scriptstyle z_L}$}
\psfrag{z1}{$z_1$}
\psfrag{z2}{$z_2$}
\psfrag{zL}{$z_L$}
\psfrag{post_prob_decoupled}{${\scriptstyle P(\bmx|\bmz)}$}
\psfrag{PLbL}{${\scriptstyle \tilde{P}^L(\bmx^L)}$}
\begin{center}
\includegraphics[scale=1.0]{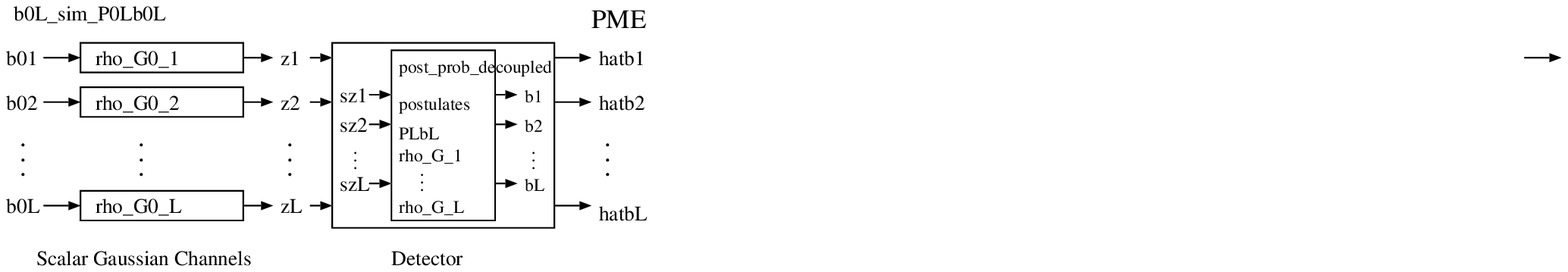}
(b) \vspace{-3mm}
\end{center}
\caption{The linear vector channel and the corresponding detector (a).
The bank of scalar Gaussian channels and their corresponding detector (b).
}
\label{fig:channel}
\end{figure}

Detailed derivation of the claim is given 
in Section~\ref{sec:derivation}.  
The claim implies that 
the scalar-channel characterization is valid 
for the joint distribution $\mathcal{P}(\bmx_0^L,\,\bmx^L)$, 
this is, 
the joint distribution $\mathcal{P}(\bmx_0^L,\,\bmx^L)$
defined in \eqref{eq:joint_distt} 
can be asymptotically
identified as the joint distribution of $\bmx_0^L$ and 
$\bmx^L$ where the elements of $\bmx_0^L$ are
independently transmitted over the scalar Gaussian channel $\rho_{G0}(z|x)$ 
and where the detector postulates the channel model $\rho_G(z|x)$ and the
modulated version of the assumed prior $\tilde{P}(\bmx^L)$ (Fig.
\ref{fig:channel}). 
This result is a finer version of the decoupling principle,
which is first stated by Tse and Hanly~\cite{ART__DT_SVH_IEEE_Trans_IT_1999},
and named by Guo and Verd\'{u}~\cite{ART__DG_SV_IEEE_Trans_IT_2005}.

\section{Derivation of the claim}
\label{sec:derivation}

\subsection{Replica method}
We evaluate $\mathcal{P}(\bmx_0^L,\,\bmx^L)$ 
defined in~\eqref{eq:joint_distt} via replica method.  
Introducing a real number $n$,
\eqref{eq:joint_distt} can be rewritten as
\begin{align}
&\mathcal{P}(\bmx_0^L,\,\bmx^L)
= \!\!\! \lim_{K,\,N\to\infty}\lim_{n\to0}\EE_H
\Bigg[
\! \int \! \left\{
\int P(\bmy|H \bmx) P(\bmx) \,d\bmx^{\backslash L}
\right\}
\nonumber \\
&\times
\left\{
\int
\!\!
P(\bmy|H \bmx) P(\bmx) d\bmx
\right\}^{n-1}
\!\!\!\!\!
P_0(\bmy|H \bmx_0)P_0(\bmx_0) d\bmx_0^{\backslash L}
\Bigg].
\end{align}
According to the standard prescription of replica method, 
we first evaluate
\begin{align}
\mathcal{Z}_n(\bmx_0^L,\,\bmx_1^L) 
&= \lim_{K,\,N \to \infty}
\EE_H \Bigg[
\prod_{a=0}^n \Bigg\{
\int
P_a(\bmy|H \bmx_a) P_a(\bmx_a)
\Bigg\}
\nonumber \\
&\times
d\bmx_0^{\backslash L} d\bmx_1^{\backslash L}
\prod_{a=2}^n d\bmx_a
\Bigg]
\label{eq:partition_function}
\end{align}
for a positive integer $n$, 
where $P_a(\bmy|H \bmx_a) = P(\bmy|H \bmx_a)$ and $P_a(\bmx)=P(\bmx)$ for 
$a=1,\,\ldots,\,n$, 
and then the result is continuated to real $n$ 
in order to take the limit $n\to0$ to obtain 
\begin{align}
\lim_{n \to 0} \mathcal{Z}_n(\bmx_0^L,\,\bmx_1^L) \Big|_{\bmx_1^L=\bmx^L}
= \mathcal{P} (\bmx_0^L,\,\bmx^L).
\end{align}
Although there is no rigorous justification for the replica method,
we assume validity of the replica method and related techniques 
throughout this paper.

\subsection{Average over channel matrix}
To evaluate \eqref{eq:partition_function}, we first take  
the average over the channel matrix $H$.
Using the assumptions of random channel matrix and memoryless channels,
one has 
\begin{align}
&\mathcal{Z}_n(\bmx_0^L,\,\bmx_1^L)
\nonumber \\
&= \lim_{K,\,N \to \infty}
\idotsint
\left\{
\EE_{\bmh} \left[
\int \prod_{a=0}^n \rho_a
\left(y \Bigg| \frac{\bmh^T \bmx_a}{\sqrt{N}} \right)\,dy
\right]
\right\}^N
\nonumber \\
&\times
\prod_{a=0}^n P_a(\bmx_a)
\,d\bmx_0^{\backslash L} \,d\bmx_1^{\backslash L}\,
\prod_{a=2}^n d\bmx_a,
\label{eq:partition_function_no1}
\end{align}
where $\rho_a(y|u) = \rho(y|u)$ for $a=1,\,\ldots,\,n$.
We let 
\begin{align}
A = \left\{
\EE_{\bmh} \left[
\int \prod_{a=0}^n \rho_a
\left(y \Bigg|\frac{\bmh^T \bmx_a}{\sqrt{N}} \right)\,dy
\right]
\right\}^N
\label{eq:def_A}
\end{align}
and introduce auxiliary random variables $\bmv=(v_0,\,\ldots,\,v_n)^T$, 
$v_a = \bmh^T \bmx_a/\sqrt{K}$.
The average over $\bmh$ in \eqref{eq:def_A} can be rewritten 
in terms of an integral over the conditional distribution of 
$\bmv$ given $\{\bmx_a ;\,a=0\,\ldots,n\}$, 
denoted by $V(\bmv|\{\bmx_a\})$, as 
\begin{align}
A= \left\{\int V(\bmv|\{\bmx_a\}) \int \prod_{a=0}^n
 \rho_a\left(y|\sqrt{\beta}\, v_a\right)
\,dy\,d\bmv\right\}^N.
\end{align}
To obtain an explicit expression for $V(\bmv|\{\bmx_a\})$,
we evaluate the 
characteristic function
of $V(\bmv|\{\bmx_a\})$, as
\begin{align}
&
\hat{V}(\hat{\bmv}|\{\bmx_a\})
= \int
e^{i \hat{\bmv}^T \bmv}\, V(\bmv|\{\bmx_a\}) \,d\bmv
\nonumber \\
&~~~~~~~~~~~~~
= \exp \left[
- \frac{1}{2}
\hat{\bmv}^T Q \hat{\bmv}
\right]
\nonumber \\
&\times \left\{ 1 -
\frac{3-\kappa}{24K} \sum_{a, b, c, d = 0}^n W_{abcd}\,
\hat{v}_a \hat{v}_b \hat{v}_c \hat{v}_d
+ O\left(K^{-2}\right)
\right\},
\label{eq:V_hat}
\end{align}
where $\hat{\bmv} = (\hat{v}_0,\,\ldots,\,\hat{v}_n)^T$,  
where $\kappa$ is fourth-order moment of $h_{\mu k}$, and where 
$(n+1)\times(n+1)$ symmetric matrix $Q$ and fourth-order symmetric tensor $W$
are defined as
\begin{align}
Q_{ab} &= \frac{1}{K} \sum_{k=1}^K x_{ak} x_{bk}\quad (0 \le a \le b \le
 n), \\
W_{abcd} &= \frac{1}{K} \sum_{k=1}^K x_{ak} x_{bk} x_{ck} x_{dk}
~(0 \le a \le b \le c \le d \le n).
\end{align}
Note that in the above we have to evaluate $\hat{V}(\hat{\bmv}|\{\bmx_a\})$ 
up to $O(K^{-1})$ terms.  
The inverse Fourier transform yields 
\begin{align}
V(\bmv|\{\bmx_a\})
= V_G(\bmv) -\frac{1}{K} V_\Delta(\bmv) + O(K^{-2}),
\end{align}
where 
\begin{align}
V_G(\bmv) &=
\Big[ (2 \pi)^{n+1} {\rm det}\,(Q) \Big]^{-\frac{1}{2}}
 \exp \left[
-\frac{1}{2} \bmv^T Q^{-1} \bmv
\right], \\
V_\Delta(\bmv) &= \frac{3-\kappa}{24} \sum_{a,b,c,d=0}^n
W_{abcd} \,
\frac{\partial^4}{\partial v_a \partial v_b \partial v_c \partial v_d}
V_G(\bmv).
\end{align}
Collecting these expressions, we have 
\begin{align}
A
= \exp \Big[
N \mathcal{G}_0 (Q) - \mathcal{G}_1(Q,W) + O\left(K^{-1}\right)
\Big],
\label{eq:A_evaluated}
\end{align}
where
\begin{align}
\mathcal{G}_0 (Q) &= \log \int V_G(\bmv) \int \prod_{a=0}^n
\rho_a \left(y|\sqrt{\beta}\,v_a \right)\,dy\,d\bmv,
\\
\mathcal{G}_1 (Q,\,W) &=
\frac{\int V_\Delta(\bmv)
\int \prod_{a=0}^n \rho_a \left(y|\sqrt{\beta}\,v_a \right)\,dy\, d\bmv}
{\beta \int V_G(\bmv)
\int \prod_{a=0}^n \rho_a \left(y|\sqrt{\beta}\,v_a \right)\,dy\, d\bmv}.
\end{align}

\subsection{Integral over $Q$ and $W$}
Since the quantity $A$ depends on $\{\bmx_a\}$ only through $Q$ and $W$, 
one can rewrite~\eqref{eq:partition_function_no1}
in terms of an integral over $Q$ and $W$, as 
\begin{align}
&\mathcal{Z}_n(\bmx_0^L,\,\bmx_1^L)
\nonumber \\
&=
\lim_{K,\,N \to \infty} 
\iint \exp\Big[
N \mathcal{G}_0(Q) - \mathcal{G}_1(Q,\,W) + O\left(K^{-1} \right)
\Big]\,
\nonumber \\
&
\times \mu_K(Q,\,W;\,\bmx_0^L,\, \bmx_1^L) \,dQ\,dW, 
\label{eq:Z_after_introducing_order_parameter}
\end{align}
where 
\begin{align}
&\mu_K(Q,\,W;\,\bmx_0^L,\,\bmx_1^L)
\nonumber \\
&= \idotsint
\prod_{0 \le a \le b \le n}
\delta \left( \! Q_{ab} -  \frac{1}{K}\sum_{k=1}^K x_{ak} x_{bk} \! \right)
\nonumber \\
&\times
\prod_{0 \le a \le b \le c \le d \le n}
 \delta\left(W_{abcd}-\frac{1}{K}\sum_{k=1}^K x_{ak} x_{bk} x_{ck} x_{dk} \right)
\nonumber \\
&\times
\prod_{a=0}^n P_a (\bmx_a)
\,d\bmx_0^{\backslash L} \,d\bmx_1^{\backslash L}
\,\prod_{a=2}^n d\bmx_a,
\label{eq:QWmeasure}
\end{align}
and $dQ = \prod_{0 \le a \le b \le n} dQ_{ab}$ , 
$dW = \prod_{0 \le a \le b \le c \le d \le n} dW_{abcd}$.

We evaluate \eqref{eq:QWmeasure} in the large-system limit 
by following the derivation in~\cite{ART__AML1982,ART__EB_1986}. 
We introduce parameters 
$\hat{Q}=\{\hat{Q}_{ab} ;\,0 \le a \le b \le n\}$ and 
$\hat{W}=\{\hat{W}_{abcd} ;\,0 \le a \le b \le c \le d \le n\}$, 
which are conjugates to $Q$ and $W$, respectively, 
and define some functions of them for later use: 
\begin{align}
&\Lambda (\hat{Q},\,\hat{W})
= \frac{1}{K} \log
\idotsint \prod_{k=1}^K \exp \Bigg[
\sum_{0 \le a \le b \le n} \!\!\! \hat{Q}_{ab} \, x_{ak} x_{bk}
\nonumber \\
&+ \!\!\!\!\!
\sum_{0 \le a \le b \le c \le d \le n}
\!\!\!\!\!\!\!\!\!
\hat{W}_{abcd} \,
 x_{ak} x_{bk} x_{ck} x_{dk} \Bigg]
\prod_{a=0}^n \bigg\{ P_a(\bmx_a)\,d\bmx_a \bigg\},
\label{eq:Lambda}
\end{align}
\begin{align}
&\lambda_x(\hat{Q},\, \hat{W};\bmx_0^L,\,\bmx_1^L)
\nonumber \\
&= \log
\idotsint \!
\prod_{k=1}^L
\exp\Bigg[
\sum_{0 \le a \le b \le n} \!\!\! \hat{Q}_{ab}\, x_{ak} x_{bk}
\nonumber \\
&+ \!\!\!\!\!
\sum_{0\le a \le b \le c \le d \le n}
\!\!\!\!\!\!\!\!\!
\hat{W}_{abcd}\, x_{ak} x_{bk}
x_{ck} x_{dk} \Bigg]
\prod_{a=0}^n P_a^L(\bmx_a^L)
\,\prod_{a=2}^n d\bmx_a^L,
\label{eq:lambda_b}
\end{align}
\begin{align}
&\lambda(\hat{Q},\,\hat{W})=
\log
\idotsint
\prod_{k=1}^L
\exp\Bigg[
\sum_{0 \le a \le b \le n} \hat{Q}_{ab}\, x_{ak} x_{bk}
\nonumber \\
&+
\!\!\!\!\!
\sum_{0\le a \le b \le c \le d \le n}
\!\!\!\!\!\!\!\!\!
 \hat{W}_{abcd}\, x_{ak} x_{bk}
x_{ck} x_{dk} \Bigg]
\prod_{a=0}^n \bigg\{P_a^L (\bmx_a^L)\,d\bmx_a^L\bigg\}.
\label{eq:lambda}
\end{align}
We further assume that $\Lambda(\hat{Q},\,\hat{W})$ has a limit as $K
\to \infty$.
Using the functions \eqref{eq:Lambda}--\eqref{eq:lambda},
the Fourier transform of \eqref{eq:QWmeasure} is 
given by 
\begin{align}
&
\hat{\mu}_K (\hat{Q},\,\hat{W};\,\bmx_0^L,\,\,\bmx_1^L )
=\exp \Bigg[
K \Lambda \left(i \frac{\hat{Q}}{K},\,i \frac{\hat{W}}{K}\right)
\nonumber \\
&+ \lambda_x \left(i \frac{\hat{Q}}{K},\,i
\frac{\hat{W}}{K};\,\bmx_0^L,\,\bmx_1^L \right)
- \lambda \left(i \frac{\hat{Q}}{K},\,i\frac{\hat{W}}{K} \right)
\Bigg], 
\end{align}
and its inverse Fourier transform yields
\begin{align}
&\mu_K(Q,\,W;\,\bmx_0^L,\,\bmx_1^L)
=
\left( \frac{K}{2\pi} \right)^{
\left\{
\genfrac{(}{)}{0pt}{}{n+2}{2}+
\genfrac{(}{)}{0pt}{}{n+4}{4}
\right\}
}
\nonumber \\
&\times
\iint
\exp\left[
K \left\{
-i Q \cdot \hat{Q}
- i W \cdot \hat{W}
+ \Lambda(i\hat{Q},\,i\hat{W})
 \right\}
\right]
\nonumber \\
&\times \exp \left[
\lambda_x(i\hat{Q},\,i\hat{W};\,\bmx_0^L,\,\bmx_1^L) -
\lambda(i\hat{Q},\,i\hat{W})
\right]
\,d\hat{Q}\,d\hat{W},
\label{eq:mu_with_QWhat}
\end{align}
where $Q \cdot \hat{Q}$ and $W \cdot \hat{W}$ are
abbreviations of
$\sum_{0 \le a \le b \le n} Q_{ab}\, \hat{Q}_{ab}$ and
$\sum_{0 \le a \le b \le c \le d \le n} W_{abcd}\, \hat{W}_{abcd}$, 
respectively.

To evaluate the integral over $\hat{Q}$ and $\hat{W}$
in \eqref{eq:mu_with_QWhat}, 
let 
$\hat{Q}^*=\{\hat{Q}_{ab}^*;\,0 \le a \le b \le n\}$ and 
$\hat{W}^*=\{\hat{W}_{abcd}^*;\,0 \le a \le b \le c \le d \le n \}$
denote the solution of the equations 
\begin{align}
Q_{ab} = \frac{\partial \Lambda(\hat{Q},\,\hat{W})}{\partial \hat{Q}_{ab}}
,~
W_{abcd} = \frac{\partial \Lambda(\hat{Q},\,\hat{W})}{\partial \hat{W}_{abcd}}.
\label{eq:fixed_point_eq_QWhat}
\end{align}
Applying three operations to \eqref{eq:mu_with_QWhat}; 
a change of variables 
\begin{align}
i \hat{Q}_{ab} \rightarrow i \frac{\hat{Q}_{ab}}{\sqrt{K}} + \hat{Q}^*_{ab},~
i \hat{W}_{abcd} \rightarrow i \frac{\hat{W}_{abcd}}{\sqrt{K}} +
\hat{W}^*_{abcd}, 
\end{align}
Taylor expansion of $\Lambda$, $\lambda_x$ and $\lambda$,
and a change of integration paths to real axes,
one can find that the integral in
\eqref{eq:mu_with_QWhat} leads to a Gaussian integration. 
Then, one obtains
\begin{align}
&\mu_K(Q,\,W;\,\bmx_0^L,\,\bmx_1^L)
\nonumber \\
&=
\left( \frac{K}{2\pi} \right)^{
\frac{1}{2}\left\{
\genfrac{(}{)}{0pt}{}{n+2}{2}+
\genfrac{(}{)}{0pt}{}{n+4}{4}\right\}}
{\rm det}\left( \mathcal{H}
(\Lambda|\hat{Q}^*,\,\hat{W}^*)\right)^{-\frac{1}{2}} \nonumber \\ &\times
\exp\Big[
K \left\{
 - Q \cdot \hat{Q}^* - W \cdot \hat{W}^*
 + \Lambda(\hat{Q}^*,\,\hat{W}^*)
 \right\}
\nonumber \\
&+
\lambda_x(\hat{Q}^*,\,\hat{W}^*;\,\bmx_0^L,\,\bmx_1^L) -\lambda(\hat{Q}^*,\,\hat{W}^*)
+ O\left(K^{-\frac{1}{2}}\right)
\Big], 
\label{eq:measure}
\end{align}
where $\mathcal{H}(f|\bm{u}^*)$ represents 
a Hessian matrix of the function $f(\bm{u})$ at $\bm{u}=\bm{u}^*$.
Use of Gaussian integration requires 
the Hessian
matrix $\mathcal{H}(\Lambda|\hat{Q}^*,\,\hat{W}^*)$ being 
positive definite.
Note that a similar evaluation is still possible 
when $\mathcal{H}(\Lambda|\hat{Q}^*,\,\hat{W}^*)$ is 
non-negative definite~\cite{ART__EB_1987}.

\subsection{Saddle-point evaluation}
We evaluate the integral over $Q$ and $W$ in
\eqref{eq:Z_after_introducing_order_parameter} via the saddle-point
method~\cite{BOO_ETC_1965}.
We obtain 
\begin{align}
&\mathcal{Z}_n(\bmx_0^L,\,\bmx_1^L)
\nonumber \\
&= \lim_{K,\,N\to\infty}
D
\,\exp \Big[
K n \mathcal{F}_n(Q^*,\,W^*)
- \mathcal{G}_1(Q^*,\,W^*)
\nonumber \\
&+ \lambda_x(\hat{Q}^*,\,\hat{W}^*;\,\bmx_0^L,\,\bmx_1^L)
- \lambda(\hat{Q}^*,\,\hat{W}^*)
+ O\left( K^{-1} \right)
\Big],
\label{eq:Z_QW}
\end{align}
where the function $\mathcal{F}_n(Q,\,W)$
is defined as 
\begin{align}
&\mathcal{F}_n(Q,\,W)
\nonumber \\
&= \frac{1}{n} \left[
\frac{1}{\beta}
\mathcal{G}_0(Q) - Q \cdot \hat{Q}^* -W \cdot \hat{W}^* +
\Lambda(\hat{Q}^*,\,\hat{W}^*)
\right]
\label{eq:free_energy_n}
\end{align}
Note that $\hat{Q}^*$ and $\hat{W}^*$ depend on $Q$ and $W$ via
\eqref{eq:fixed_point_eq_QWhat}.
The saddle points $Q^*=\{Q_{ab}^*;\,0 \le a \le b \le n\}$
and $W^* = \{W_{abcd}^*;\,0 \le a \le b \le c \le d \le n\}$
are determined as the solution of
\begin{align}
\frac{\partial \mathcal{F}_n(Q,\,W)}{\partial Q_{ab}} = 0,\quad
\frac{\partial \mathcal{F}_n(Q,\,W)}{\partial W_{abcd}} = 0.
\label{eq:fixed_point_eq_QW}
\end{align}
If more than one solution exists for \eqref{eq:fixed_point_eq_QW}, 
the correct solution is the one that maximizes~\eqref{eq:free_energy_n}.
The normalization factor $D$ is given by 
\begin{align}
D
=
\left[ {\rm det}\Big(
\mathcal{H}(\Lambda|\hat{Q}^*,\,\hat{W}^*)
\Big)
{\rm det}\Big(
\mathcal{H}(-n \mathcal{F}_n|Q^*,\,W^*)
\Big)\right]^{-\frac{1}{2}}.
\end{align}
Application of the saddle-point method here requires that 
the Hessian matrix $\mathcal{H}(-n \mathcal{F}_n|Q^*,\,W^*)$
is positive definite. 

Since our final result will be a function of $\bmx_0^L$ and $\bmx_1^L$, 
we can ignore terms in \eqref{eq:Z_QW} which are independent of 
these variables, obtaining 
\begin{align}
&\mathcal{Z}_n (\bmx_0^L,\,\bmx_1^L) \propto
\exp\left[
\lambda_x(\hat{Q}^*,\,0;\,\bmx_0^L,\,\bmx_1^L)
\right].
\label{eq:partition_function_n_b}
\end{align}
Note that one obtains $\hat{W}_{abcd} = 0$ by solving
\eqref{eq:fixed_point_eq_QW}, and that 
the overall factor, which we have just ignored, can be
determined via normalization.
It turns out, from $\hat{W}_{abcd} = 0$, 
\eqref{eq:fixed_point_eq_QWhat}, and \eqref{eq:fixed_point_eq_QW}, 
that $Q^*$ and $\hat{Q}^*$ do not depend on $W^*$.

\subsection{Replica symmetric ansatz}
To proceed further, we assume replica symmetry (RS)~\cite{INP__TT_ALT_2004},
under which we let 
\begin{align}
Q_{00}^* &= r_0, &
Q_{aa}^* &= r, &
Q_{0a}^* &= m, &
Q_{ab}^* &= q,
\nonumber \\
\hat{Q}_{00}^* &= \frac{1}{2} G_0, &
\hat{Q}_{aa}^* &= \frac{1}{2} G, &
\hat{Q}_{0a}^* &= E, &
\hat{Q}_{ab}^* &= F,
\label{eq:RS_ansatz}
\end{align}
for positive integers $a < b$.  
Then, $\mathcal{F} \equiv \lim_{n \to 0} \mathcal{F}_n(Q,\,W)$ is reduced 
to \eqref{eq:free_energy}, and the saddle-point equations 
\eqref{eq:fixed_point_eq_QWhat} and 
\eqref{eq:fixed_point_eq_QW} become 
\eqref{eq:SaddlePoint_G}--\eqref{eq:SaddlePoint_q}, \eqref{eq:SaddlePoint_r0}
and $G_0=0$ (For detailed derivation, see~\cite{INP__TT_ALT_2004}).
Notice that the condition for the Hessian matrix  
$\mathcal{H}(-n \mathcal{F}_n|Q^*,\,W^*)$ being positive definite 
yields the de Almeida-Thouless (AT) condition for local stability
of RS solutions~\cite{ART__JRLDA_DJT_1978}.

Inserting the RS assumption \eqref{eq:RS_ansatz} into
\eqref{eq:partition_function_n_b}, one obtains 
\begin{align}
&\mathcal{Z}_n (\bmx_0^L,\,\bmx_1^L)
\nonumber \\
&\propto \int
\left[
\prod_{k=1}^L
 \rho_G(z_k|x_{1k})
e^{\frac{G-F+E}{2} \Ltwo \bmx_1^L \Ltwo^2} P_1^L(\bmx_1^L)
\right]
\nonumber \\
&
\times \left[
\int \prod_{k=1}^L
\rho_G(z_k|x_k)
e^{\frac{G-F+E}{2} \Ltwo \bmx^L \Ltwo^2} P^L(\bmx^L)
\,d\bmx^L\right]^{n-1}
\nonumber \\
&\times
\prod_{k=1}^L\left\{
\rho_{G0}(z_k|x_{0k}) e^{\frac{1}{2} \left( n E z_k^2 + G_0 x_{0k}^2 \right)}
\right\} P_0^L(\bmx_0^L) \,d\bmz^L.
\label{eq:partition_func_n_b_RS}
\end{align}
Taking the limit $n\to 0$, 
one finally arrives at \eqref{eq:decoupled_joint_distt}.

\section{Conclusion}
In this paper, we have considered the decoupling principle of 
the linear vector channel.
We have shown that 
the scalar-channel characterization obtained via decoupling principle
is valid
for the joint distributions of $L\,(\ll K)$ elements of input vector
and their estimates based on the posterior probability,
in the large-system limit.
This implies that the scalar-channel characterization
is valid not only for macroscopic quantities, 
but also for microscopic quantities
on the linear vector channel.

\section*{Acknowledgment}
The authors would like to acknowledge support from 
the Grant-in-Aid for Scientific Research on Priority Areas (No.~18079010), 
the Ministry of Education, Culture, Sports, Science and Technology,
Japan.

\bibliographystyle{IEEEtran}
\bibliography{IEEEabrv,kn_isit2008}

\end{document}